\documentclass{appolb}
\usepackage{graphicx}
\usepackage{amsmath}
\usepackage{graphics,bm}
\usepackage{epsfig}
\usepackage{graphicx}
\usepackage{amsmath}
\usepackage{wrapfig}
\usepackage{color}
\usepackage{caption}
\usepackage{hyperref}
\usepackage[toc,page]{appendix}
\newcommand{\sign}{\textrm{sgn}}

\begin{document}
\title{Vortex solutions in the Abelian Higgs Model with a neutral scalar
%\thanks{Presented at ...}%
}
\author{Prabal Adhikari, Jaehong Choi
\address{St. Olaf College, 1520 St. Olaf Avenue, Northfield, MN 55057}
%\\
%{Third Author of different affiliation
%}
%the Name(s) of other Author(s)
%\address{affiliation}
}
\maketitle
\begin{abstract}
We construct an extension of the Abelian Higgs model, which consists of a complex scalar field by including an additional real, electromagnetically neutral scalar field. We couple this real scalar field to the complex scalar field via a quartic coupling and investigate $U(1)$ vortex solutions in this ``extended Abelian Higgs Model". Since this model has two additional homogeneous ground states, the $U(1)$ vortices that can form in this model have a richer structure than in the Abelian Higgs Model. We also find the ``phase diagram" of the model showing the parameter space in which the real scalar particle condenses in the vortex state while having a zero vacuum expectation value in the homogeneous ground state. 
\end{abstract}
%\PACS{PACS numbers come here}
  
\section{Introduction}
Magnetic $U(1)$ vortices have been of great interest both theoretically and experimentally since they were first predicted by Abrikosov~\cite{Abrikosov} in Ginzburg-Landau theory~\cite{LandauGinzburg} back in the fifties. Since then they have been studied extensively not only in the context of non-relativistic condensed matter systems~\cite{Leggett} but also in particle physics~\cite{Nielsen,Nielsen2,Olesen,Vega}. It is interesting to note that a mathematical isomorphism between Ginzburg-Landau theory and the Abelian Higgs model was formally established~\cite{Harrington}  in the seventies. More recently, there has been increasing interest in the properties of Quantum Chromodynamics (QCD) in the presence of large magnetic fields~\cite{Jens}. This interest stems from the presence of large magnetic fields in a variety of physical systems including magnetars~\cite{Lai} and in quark gluon plasma~\cite{CME1,CME2,CME3} at RHIC.

Additionally, there has also been increasing interest in systems of mesons in the presence of magnetic fields. It was shown that $\rho$-mesons may condense in the QCD vacuum at extremely large magnetic fields~\cite{Maxim,Maxim2,Maxim3}. The resulting condensed system consists of charged $\rho^{\pm}$ mesons and a neutral $\rho^{0}$ meson and in the presence of magnetic field they can form a ground state consisting of a magnetic vortex lattice, in which not only the charged $\rho$-mesons condense but also the neutral meson superfluid condenses. On the other hand, there are systems such as low-energy QCD described by chiral perturbation theory, which at finite isospin and in the presence of external magnetic fields behaves like a type-II superconductor. Chiral perturbation theory has three mesons, namely the charged pions $\pi^{\pm}$ and the neutral pion, $\pi^{0}$. However, in the magnetic vortices that form at moderate external fields only the charged pions ($\pi^{\pm}$) condense in the vortex state and the neutral pions are absent in the vortices~\cite{Julia}. This is also true in the context of the linear sigma model~\cite{Harrington2}.

The physics encoded within chiral perturbation theory is constrained by the symmetries and symmetry breaking properties of QCD. In this paper, we consider a less restrictive model: an extension of the Abelian Higgs model consisting of charged scalars coupled to a charge-less scalar via a quartic coupling and investigate vortex solutions within this model. In particular we are interested in studying the possibility of neutral scalar condensation. Of course there is the trivial scenario, where the real scalar field has a non-zero vacuum expectation value (vev) and as condenses in a vortex. But a slightly more non-trivial scenario is when the real scalar field has a zero vacuum expectation value, yet condenses in the vortex phase. The advantage of using such a model is that there are parameters such as the scalar masses and strength of the interaction between the scalars that can be controlled less restrictively.

The paper is organized as follows: in section 2, we construct the ``extended Abelian Higgs model" and find the homogeneous ground state of the theory. In section 3, we consider the behavior of the vortex asymptotically far away from the vortex center. In section 4, we show the full numerical vortex solutions for a wide range of parameters and interactions between the charged and the neutral scalars. Finally, we construct a ``phase diagram" showing the region in parameter space, where neutral pion condenses while having a zero vacuum expectation value.

\section{The extended Abelian Higgs Model} 
We begin by constructing the Lagrangian for an extended version of the standard Abelian Higgs model that not only contains a set of charged scalars ($\Phi$, $\Phi^{\dagger}$) but also a real scalar field ($\phi$) that couples to the complex scalar field via an interaction term of the form $f|\Phi|^{2}|\phi|^{2}$, where $f$ is the strength of the coupling between the real and complex scalar fields. Note that the two fields cannot couple via electromagnetic interactions since the real scalar is electromagnetically neutral. Therefore, the fields must couple via strong interactions - it is worth pointing out that in chiral perturbation, the two charged pions ($\pi^{\pm}$) and the neutral pion ($\pi^{0}$) interact with each other through both momentum-dependent and momentum-independent interactions~\cite{CPT}.

The effective Lagrangian for the extended Abelian Higgs Model is then:
\begin{equation}
\label{Lagrangian}
\mathcal{L}=-\frac{1}{4}F_{ij}F^{ij}-|(\vec{\nabla}+ie\vec{A})\Phi|^{2}-\frac{1}{2}(\vec{\nabla}\phi)^{2}-V(|\Phi|^{2}, |\phi|^{2}) \ ,
\end{equation}
where
\begin{equation}
\label{potential}
V(|\Phi|^{2}, |\phi|^{2})=a |\Phi|^{2}+b|\Phi|^{4}+c|\phi|^{2} +d|\phi|^{4}+f|\Phi|^{2} |\phi|^{2}\ .
\end{equation}
In the Lagrangian, $F_{ij}\equiv\partial_{i}A_{j}-\partial_{i}A_{j}$ is the electromagnetic tensor with $A_{i}$ representing the spatial components of the vector potential. $\Phi$ is a complex scalar field and $\phi$ is a real scalar field, which is electromagnetically neutral. $\pm e$ is the charge associated with the complex scalar fields. Finally, note that the Lagrangian is invariant under local U(1) gauge transformations, i.e. 
\begin{equation}
\begin{split}
\Phi&\rightarrow e^{i\Lambda(\vec{x})}\Phi\\
e\vec{A}&\rightarrow e\vec{A}+\vec{\nabla}\Lambda(\vec{x})\ ,
\end{split}
\end{equation}
and under $|\phi|\rightarrow -|\phi|$ and $|\Phi|\rightarrow -|\Phi|$.
%%%%%
\subsection{Ground State in the absence of photons}
In this section, we find the ground state of the model in the absence of a photon field, i.e. $\vec{A}=0$. The ground state must be homogeneous, therefore we begin by minimizing the potential. In order to do so, we introduce the notation:
\begin{equation}
X\equiv |\Phi|^{2},\ Y\equiv |\phi|^{2}\ .
\end{equation}
and minimize the potential with respect to $X$ and $Y$. We get
%\begin{equation}
%\begin{split}
%\frac{\partial V}{\partial |\Phi|}&=2|\Phi|\frac{\partial V}{\partial X}=0\\
%\frac{\partial V}{\partial |\phi|}&=2|\phi|\frac{\partial V}{\partial Y}=0\ .
%\end{split}
%\end{equation}
\begin{equation}
\begin{split}
\frac{\partial V}{\partial X}&=a+2bX+fY=0\\
\frac{\partial V}{\partial Y}&=c+2dY+fX=0\ .
\end{split}
\end{equation}
The extremum of the potential then occurs when
\begin{equation}
\label{gs}
%v_{\Phi}=\sqrt{\frac{cf-2ad}{4bd-f^{2}}}\textrm{ and }v_{\phi}=\sqrt{\frac{af-2bc}{4bd-f^{2}}}\ ,
X_{\rm ext}=\frac{cf-2ad}{4bd-f^{2}}\textrm{ and }Y_{\rm ext}=\frac{af-2bc}{4bd-f^{2}}\ ,
\end{equation}
where $X_{\rm ext}$ and $Y_{\rm ext}$ are the magnitude squared of the fields $\Phi$ and $\phi$ (respectively) at the extremum of the potential $V$ in Eq.~(\ref{potential}). The extremum is a locally stable minimum if the following conditions are satisfied (according to the second derivative test):
\begin{equation}
\label{first derivative}
%\begin{split}
\frac{\partial^{2} V}{\partial X^{2}}=2b>0\\
%\frac{\partial^{2} V}{\partial Y^{2}}&=2d\\
%\frac{\partial^{2} V}{\partial X\partial Y}&=\frac{\partial^{2} V}{\partial Y\partial X}=f\ .
%\end{split}
\end{equation}
and
\begin{equation}
\label{dc}
D\equiv\frac{\partial^{2} V}{\partial X^{2}}\frac{\partial^{2} V}{\partial Y^{2}}-\left(\frac{\partial^{2} V}{\partial X\partial Y}\right)^{2}=4bd-f^{2}>0\ ,
\end{equation}
which automatically implies that $d\ge 0$ since $f^{2}$ is positive definite.

Since by definition, $X_{\rm min}\ge 0$ and $Y_{\rm min}\ge 0$ and the local minimum stability condition requires that the discriminant, $D\equiv 4bd-f^{2}>0$, it follows that the model possesses the following vacuum expectation values in their ground states. The vevs of $\Phi$ and $\phi$ are denoted by $v_{\Phi}$ and $v_{\phi}$ respectively:\\
\\
1. normal vacuum:\\
\begin{equation}
v_{\Phi}=v_{\phi}=0 \textrm{ if }cf\le 2ad\textrm{ and }af\le 2bc
\end{equation}
\\
2. charged superfluid:\\
\begin{equation}
v_{\Phi}=\sqrt{\frac{-a}{2b}},\ v_{\phi}=0 \textrm{ if }cf>2ad\textrm{ and }af\le 2bc
\end{equation}
\\
3. neutral superfluid:\\
\begin{equation}
v_{\Phi}=0,\ v_{\phi}=\sqrt{\frac{-c}{2d}}\textrm{ if }cf\le 2ad\textrm{ and }af>2bc
\end{equation}
\\
4. charged and neutral superfluids\\
\begin{equation}
v_{\Phi}=\sqrt{\frac{cf-2ad}{4bd-f^{2}}},\ v_{\phi}=\sqrt{\frac{af-2bc}{4bd-f^{2}}}\textrm{if }cf>2ad\textrm{ and }af>2bc\ .
\end{equation}
\\
Note that the vacuum expectation values in the normal and charged superfluid state are consistent with the Abelian Higgs Model~\cite{Nielsen2,Olesen} as expected. Also, local stability requires $D>0$ and $b>0$ (and consequently $d>0$).

\section{Vortex solutions} 
In this section, we consider single vortex solutions within the extended Abelian Higgs model. We assume time-independence, i.e. $\partial_{0}A_{i}=0$ and that the zeroth component of the vector potential, $A_{0}=0$. In other words, we will only concern ourselves with magnetic fields in this paper and ignore electric fields. 

It is important to note that when considering the thermodynamics one has to be careful not to ignore the energy density associated with the electromagnetic charge densities that condense in the charged superfluid state. It can be ignored as long as the electromagnetic energy density associated with the charged superfluids is small compared to the strength of the remaining interactions, see Ref.~\cite{Julia} for a full discussion in the context of chiral perturbation theory. In this paper, we will only consider single vortex solutions and leave the thermodynamic discussion to future work.
\subsection{Vortex Ansatz}
Vortices are known to be the ground state in the presence of ``moderate" external magnetic fields in Ginzburg Landau theory~\cite{Abrikosov} (and therefore the Abelian Higgs model~\cite{Harrington}). Vortex solutions are cylindrically symmetric and obey the following ansatz:
\begin{equation}
\begin{split}
A_{r}(\vec{x})&=A_{z}(\vec{x})=0,\ A_{\theta}(\vec{x})=A(r)\\
\Phi(\vec{x})&=|\Phi(r)|\exp(i\chi)\text{ where } \chi= n \theta\\
\phi(\vec{x})&=\phi(r)\ ,
\end{split}
\end{equation}
where $\theta$ is the polar angle and $n$ is an integer. Note that only the radial and longitudinal components of the vector potential are zero and the polar component only depends on the radial distance from the center of the vortex. The polar dependence of the the charged scalar field is what gives rise to the vortex quantization condition.
\subsection{Vortex Quantization Condition}
The electromagnetic current is as follows:
\begin{equation}
j_{i}=ie(\Phi^{\dagger}\partial_{i}\Phi-\Phi\partial_{i}\Phi^{\dagger})+2e^{2}A_{i}|\Phi|^{2}\ . 
\end{equation}
(Note that it is proportional to the electromagnetic charge of the complex scalar as expected.)
Using the vortex ansatz, it is straightforward to show that the radial and longitudinal components of the current are both zero. 
\begin{equation}
j_{z}=0,\ j_{r}=ie(\Phi^{\dagger}\partial_{r}\Phi-\Phi\partial_{r}\Phi^{\dagger})=0\ . 
\end{equation}
\begin{equation}
\begin{split}
j_{\theta}&=ie\left(\Phi^{\dagger}\frac{1}{r}\partial_{\theta}\Phi-\Phi\frac{1}{r}\partial_{\theta}\Phi^{\dagger}\right)+2e^{2}A(r)|\Phi|^{2}\\
&=-\frac{2e|\Phi(r)|^{2}\partial_{\theta}\chi}{r}+2e^{2}A(r)|\Phi(r)|^{2}\\
%&=2e^{2}|\Phi(r)|^{2}\left (A(r)-\frac{\partial_{\theta}\chi}{er} \right )\\
&=2e^{2}|\Phi(r)|^{2}\left (A(r)-\frac{2\pi n}{er} \right )\ .
\end{split}
\end{equation}
The polar component of the electromagnetic current is non-zero but must vanish at infinity (i.e. $r\rightarrow \infty$) giving the magnetic flux quantization condition, which is identical to the Abelian Higgs model since the new scalar field is charge-less and hence does not contribute to the electromagnetic current~\cite{Nielsen}.
\begin{equation}
\label{flux}
\int_{\textrm{C}_\infty}\vec{A}\cdot{d\vec{l}}=\int\vec{B}\cdot d\vec{a}=\frac{2n\pi}{e}\ ,
\end{equation}
where the line integral in the first term is performed at infinity. Asymptotically, the polar component of the vector potential has the following form:
\begin{equation}
\lim_{r\rightarrow \infty}A(r)=\frac{n}{er}\ ,
\end{equation}
which is consistent with the flux quantization condition of Eq.~(\ref{flux}).
\subsection{Vortex Asymptotics}
In order to analyze the vortex asymptotics, we begin with the equations of motion
\begin{equation}
\begin{split}
-\partial_{i}F_{ik}&=j_{k}\\
(\partial_{i}+ieA_{i})^{2}|\Phi|+2a|\Phi|+4b|\Phi|^{3}+2f|\Phi||\phi|^{2}&=0\\
\partial_{i}\partial_{i}|\phi|+2c|\phi|+4d|\phi|^{3}+2f|\Phi|^{2}|\phi|&=0\ ,
\end{split}
\end{equation}
which can be simplified in cylindrical coordinates using the vortex ansatz:
\begin{equation}
-\frac{\partial^{2}A}{\partial^{2}r}-\frac{1}{r}\frac{\partial A}{\partial r}+\frac{A(r)}{r^{2}}=j_{\theta}=2e^{2}|\Phi(r)|^{2}\left (A(r)-\frac{n}{er} \right )
\label{eom1}
\end{equation}
\begin{equation}
\begin{split}
-\frac{\partial^{2}|\Phi|}{\partial^{2}r}-\frac{1}{r}\frac{\partial|\Phi|}{\partial r}+\left (\frac{n}{r}-eA(r)\right )^{2}|\Phi|+2a|\Phi|+4b|\Phi|^{3}+2f|\phi|^{2}|\Phi|=0
\end{split}
\label{eom2}
\end{equation}
\begin{equation}
-\frac{\partial^{2} |\phi|}{\partial^{2} r}-\frac{1}{r}\frac{\partial |\phi|}{\partial r}+2c|\phi|+4d|\phi|^{3}+2f|\Phi|^{2}|\phi|=0
\label{eom3}
\end{equation}
While the equations above cannot be solved exactly, it is possible to deduce the behavior close to the center of the vortex ($r\rightarrow 0$):
\begin{equation}
\begin{split}
\lim_{r\rightarrow 0} A(r)&\sim r^{2n}\textrm{ with }n\ge1\\
\lim_{r\rightarrow 0} |\phi(r)|&\sim r^{2n}\textrm{ with }n\ge1\\
\lim_{r\rightarrow 0} |\Phi(r)|&\sim r^{2n} \textrm{ with }n\ge1\ ,
\end{split}
\end{equation}
and also asymptotically far away from the vortex:
\begin{equation}
\begin{split}
\lim_{r\rightarrow\infty}A(r)&=\frac{n}{er}\\
\lim_{r\rightarrow\infty}|\Phi(r)|&=v_{\Phi}\\
\lim_{r\rightarrow\infty}|\phi(r)|&=v_{\phi}\ .
\end{split}
\end{equation}
Furthermore, it is possible to find how the photon and scalar fields approach their asymptotically values. We begin with $A(r)$ written in the following form:
\begin{equation}
A(r)=\frac{n}{er}+\delta A(r)\ .
\end{equation}
Assuming $|\Phi(r)|=v_{\Phi}$, the equation of motion for $A(r)$ can be solved to get:
\begin{equation}
%a(r)&=c_{2}Y_{1}(i\sqrt{2} v_{\Phi}er)+c_{1}J_{1}(i\sqrt{2} v_{\Phi}er)\\
\delta A(r)=c_{A}K_{1}(\sqrt{2} v_{\Phi}er)+c_{B}I_{1}(\sqrt{2} v_{\Phi}er)\ ,
\end{equation}
where the $c_{i}$ s are arbitrary constants and $K_{1}$ and $I_{1}$ are modified Bessel functions of the second and first kinds. We can set $c_{B}=0$ noting that the Bessel function of the second kind, $I_{1}$, has the wrong asymptotic behavior\footnote{Note that the modified Bessel functions have the following asymptotic properties:\\ 
\begin{equation*}
\begin{split}
K_{1}(mx)&=e^{-mx}\left (\sqrt{\frac{\pi}{2mx}}+\mathcal{O}\left ( \frac{1}{(mx)^{3/2}}\right ) \right )\\
I_{1}(mx)&=e^{-mx}\left (-i\sqrt{\frac{1}{2\pi m x}}+\mathcal{O}\left (\frac{1}{(mx)^{3/2}} \right ) \right )\\
&+e^{2mx}\left (\frac{1}{2\pi m x}+\mathcal{O}\left ( \frac{1}{(mx)^{3/2}}\right ) \right )
\end{split}
\end{equation*}
}. 
Therefore, the polar component of the vector potential, $A(r)$, has the following asymptotic form:
\begin{equation}
\begin{split}
A(r)&=\frac{n}{er}+c_{A}K_{1}(\sqrt{2}v_{\Phi}er)\\
&=\frac{n}{er}+\tilde{c}_{A}e^{-\sqrt{2}v_{\Phi}er}\left (\frac{1}{\sqrt{r}}+\mathcal{O}\left (\frac{1}{r^{3/2}} \right )\right )\ ,
\end{split}
\end{equation}
with $\tilde{c}_{A}\equiv\frac{c_{A}}{2^{3/4}}\sqrt{\frac{\pi}{ev_{\Phi}}}$ with $c_{A}$ being an undetermined constant.

Next, we proceed to find the asymptotic forms of both the scalar fields, $|\Phi(r)|$ and $|\phi(r)|$. In order to do so, we first define:
\begin{equation}
\begin{split}
|\Phi(r)|&=v_{\Phi}+ \delta\Phi(r)\\
|\phi(r)|&=v_{\phi}+\delta\phi(r)\ ,
\end{split}
\end{equation}
where $v_{\Phi}$ and $v_{\phi}$ are the vevs in the ground state and $\delta\Phi$ and $\delta\phi$ are the fluctuations around the vevs. Then equations of motion for $\delta\Phi(r)$ and $\delta\phi(r)$ are as follows:
\begin{equation}
\label{coupled}
\begin{split}
\delta\Phi''(r)+\frac{\delta \Phi'(r)}{r}-n_{1}\delta \Phi(r)-n_{2}\delta \phi(r)&=v_{\Phi}e^{2}\delta A(r)^{2}\\
\delta\phi''(r)+\frac{\delta \phi'(r)}{r}-n_{4}\delta \phi(r)-n_{3}\delta \Phi(r)&=0\ ,
\end{split}
\end{equation}
where the $n_{i}$s are defined as:
\begin{equation}
\label{ns}
\begin{split}
n_{1}&\equiv2a+12b v_{\Phi}^{2}+2f v_{\phi}^{2}\\
n_{2}&=n_{3}\equiv4f v_{\phi}v_{\Phi}\\
n_{4}&\equiv2c+12 dv_{\phi}^{2}+2f v_{\Phi}^{2}\ .
%n_{3}&=4f v_{\phi}v_{\Phi}\ .
\end{split}
\end{equation}
We can then rewrite the homogeneous coupled differential equation of Eq.~(\ref{coupled}) in the following form:
\begin{equation}
\label{homogeneous}
\begin{pmatrix}
  \delta\Phi''(r)+\delta \Phi'(r)/r\\
  \delta\phi''(r)+\delta \phi'(r)/r
 \end{pmatrix}=
\begin{pmatrix}
  n_{1} & n_{2} \\
  n_{3} & n_{4}
 \end{pmatrix}
 \begin{pmatrix}
 \delta\Phi(r)\\
 \delta\phi(r)
 \end{pmatrix}\ .
\end{equation}
The above matrix equation can be solved exactly; in order to do so note that the matrix with $n_{i}$s in Eq.~(\ref{homogeneous}) has the following eigenvalues and eigenvectors
\begin{equation}
\begin{split}
\lambda_{\pm}&=\frac{1}{2}\left (n_{1}+n_{4}\pm\sqrt{(n_{1}-n_{4})^{2}+4n_{2}n_{3}} \right )\\
%\lambda_{-}&=\frac{1}{2}\left (n_{1}+n_{4}+\sqrt{(n_{1}-n_{4})^{2}+4n_{2}n_{3}} \right )\\
\vec{v}_{\pm}&=\frac{1}{\sqrt{1+s_{\pm}^{2}}}\begin{pmatrix}s_{\pm}\\1\end{pmatrix}\\
%\vec{v}_{-}&=\frac{1}{\sqrt{1+s_{2}^{2}}}\begin{pmatrix}s_{2}\\1\end{pmatrix}\\
s_{\pm}&=\frac{n_{1}-n_{4}\pm\sqrt{(n_{1}-n_{4})^{2}+4n_{2}n_{3}}}{2n_{3}}\ ,
%s_{2}&=\frac{n_{1}-n_{4}+\sqrt{(n_{1}-n_{4})^{2}+4n_{2}n_{3}}}{2n_{3}}\ , 
\end{split}
\end{equation}
where $\vec{v}_{\pm}$ are the normalized eigenvectors with eigenvalues $\lambda_{\pm}$. Then we define the matrix
\begin{equation}
P=
\begin{pmatrix}
\frac{s_{-}}{\sqrt{1+s_{-}^{2}}}&\frac{s_{+}}{\sqrt{1+s_{+}^{2}}}\\
\frac{1}{\sqrt{1+s_{-}^{2}}}&\frac{1}{\sqrt{1+s_{+}^{2}}}
\end{pmatrix}\ ,
\end{equation}
and the matrix
\begin{equation}
\label{matrixN}
N=
\begin{pmatrix}
n_{1}&n_{2}\\
n_{3}&n_{4}
\end{pmatrix}\ ,
\end{equation}
such that $P^{-1}NP=D$, where
\begin{equation}
D=
\begin{pmatrix}
\lambda_{-}&0\\
0&\lambda_{+}\ .
\end{pmatrix}\ ,
\end{equation}
and
\begin{equation}
\begin{pmatrix}
\delta\tilde{\Phi}(r)\\
\delta\tilde{\phi}(r)
\end{pmatrix}
=P^{-1}
\begin{pmatrix}
\delta\Phi(r)\\
\delta\phi(r)
\end{pmatrix}
\end{equation}
with
\begin{equation}
P^{-1}=
\begin{pmatrix}
\frac{\sqrt{1+s_{-}^{2}}}{s_{-}-s_{+}}&-\frac{s_{+}\sqrt{1+s_{-}^{2}}}{s_{-}-s_{+}}\\
-\frac{\sqrt{1+s_{+}^{2}}}{s_{-}-s_{+}}&\frac{s_{-}\sqrt{1+s_{+}^{2}}}{s_{-}-s_{+}}
\end{pmatrix}\ .
\end{equation}
The coupled equations of Eq.~(\ref{homogeneous}) then decouple in the new basis:
\begin{equation}
\begin{split}
\delta\tilde\Phi''(r)+\frac{\delta\tilde\Phi'(r)}{r}=\lambda_{-}\tilde{\Phi}(r)\\
\delta\tilde\phi''(r)+\frac{\delta\tilde\phi'(r)}{r}=\lambda_{+}\tilde{\phi}(r)\ .
\end{split}
\end{equation}
The solutions of the equations then are
\begin{equation}
\begin{split}
\delta\tilde{\Phi}(r)&=c_{\Phi}K_{0}(\sqrt{\lambda_{-}}r)+d_{\Phi}I_{0}(\sqrt{\lambda_{-}}r)\\
\delta\tilde{\phi}(r)&=c_{\phi}K_{0}(\sqrt{\lambda_{+}}r)+d_{\phi}I_{0}(\sqrt{\lambda_{+}}r)\ ,
\end{split}
\end{equation}
where $K_{0}$ and $I_{0}$ are different Bessel functions of the second and first kinds respectively. As before, we set $d_{\phi}=d_{\Phi}=0$ since $I_{0}$ has the wrong asymptotic properties\footnote{Note that for asymptotically large $x$, $K_{1}(mx)$ has the same behavior as $K_{0}(mx)$ at leading order. But $I_{1}(mx)$ is different from $I_{0}(mx)$ by a a complex conjugation, i.e. $i\rightarrow -i$ at leading order.}.
Finally then we have solutions to the homogeneous coupled equation:
\begin{equation}
\begin{pmatrix}
\delta\Phi(r)\\
\delta\phi(r)
\end{pmatrix}=
P\begin{pmatrix}
\delta\tilde{\Phi}(r)\\
\delta\tilde{\phi}(r)
\end{pmatrix}
=P\begin{pmatrix}
c_{\Phi}K_{0}(\sqrt{\lambda_{-}}r)\\
c_{\phi}K_{0}(\sqrt{\lambda_{+}}r)
\end{pmatrix}\ .
\end{equation}
%%%%
The particular solutions of this equation have been calculated in Ref.~\cite{doublevortex}, we simply quote the result here:
\begin{equation}
\begin{pmatrix}
\delta\Phi_{p}(r)\\
\delta\phi_{p}(r)
\end{pmatrix}
\approx
\begin{pmatrix}
g_{\Phi}\\
g_{\phi}
\end{pmatrix}
\frac{e^{-\alpha r}}{r}
\end{equation}
with corrections of $\mathcal{O}\left (\frac{1}{r^{2}}\right )$. Plugging the ansatz into the coupled differential equation of Eq.~(\ref{coupled}), we get for $\alpha$, $g_{\Phi}$ and $g_{\phi}$:
\begin{equation}
\label{gandalpha}
\begin{split}
\alpha&=2\sqrt{2}ev_{\Phi}\\
g_{\Phi}&=\frac{\tilde{c}_{A}^{2}e^{2}v_{\Phi}\left(n_{4}-(\frac{\alpha}{2})^{2}\right)}{n_{2}n_{3}-n_{1}n_{4}+(n_{1}+n_{4})\left (\frac{\alpha}{2}\right )^{2}-\left (\frac{\alpha}{2}\right)^{4}}\\
g_{\phi}&=-\frac{\tilde{c}_{A}^{2}e^{2}n_{3}v_{\Phi}}{n_{2}n_{3}-n_{1}n_{4}+(n_{1}+n_{4})\left(\frac{\alpha}{2}\right)^{2}-\left(\frac{\alpha}{2}\right)^{4}}\ .
\end{split}
\end{equation}
Therefore, the full solution to the coupled equations of Eq.~(\ref{coupled}) is:
\begin{equation}
\label{fullsolution}
\begin{split}
\delta\Phi(r)&\approx g_{\Phi}\frac{e^{-\alpha r}}{r}+s_{-}p_{21} c_{\Phi}K_{0}(\sqrt{\lambda_{-}}r)+s_{+}p_{22} c_{\phi}K_{0}(\sqrt{\lambda_{+}}r)\\
\delta\phi(r)&\approx g_{\phi}\frac{e^{-\alpha r}}{r}+p_{21} c_{\Phi}K_{0}(\sqrt{\lambda_{-}}r)+p_{22} c_{\phi}K_{0}(\sqrt{\lambda_{+}}r)\ ,
\end{split}
\end{equation}
where $p_{ij}$ are the elements of the matrix $P$ and $g_{\Phi}$, $g_{\phi}$, $c_{\Phi}$ and $c_{\phi}$ are undetermined constants. However, note that the solution quoted in Eq.~(\ref{fullsolution}) is not valid if the vev of $\phi$ is zero. The case with $v_{\phi}=0$ is discussed towards the end of the next section.
\section{Numerical Vortex Solutions}
In this section, we present the single vortex solutions (with unit flux of $\frac{2\pi}{e}$) generated numerically. Before we proceed with a discussion of these solutions, we note that the mass of the scalars and the photon in the extended Abelian Higgs model are 
\begin{equation}
m_{\Phi}=\sqrt{a},\ m_{\phi}=\sqrt{2c}\textrm{ and }m_{\vec{A}}=ev_{\Phi}\ .
\end{equation}
It is interesting to note that the mass scales that affect the asymptotic behavior of the charged and neutral scalar condensates are $\lambda_{\pm}$ and $\alpha=2\sqrt{2}m_{\vec{A}}$. The mass scale that affects the asymptotic behavior of the vector potential is $\frac{\alpha}{2}$, not $m_{\vec{A}}$.

In our numerical vortex solutions, we use $e=1$ and $\sign(a)=-1$. Furthermore, the magnitude of $|a|$ is used to set the scale in the problem. The following dimensionless quantities are used in our numerical plots:
\begin{equation}
\overline{r}=\sqrt{|a|}r,\ \bar{\Phi}=\frac{\overline{\Phi}}{\sqrt{|a|}},\ \bar{\phi}=\frac{\overline{\phi}}{\sqrt{|a|}},\ \overline{B}=\frac{B}{|a|}\ . \\
%\bar{A_{\phi}}&=\frac{A_{\phi}}{|a|}\ .
\end{equation}
Note that $\vec{B}=\vec{\nabla}\times \vec{A}$, $B\equiv |\vec{B}|$ and that $b$, $d$ and $f$ are dimensionless parameters.
\subsection{$v_{\Phi}=v_{\phi}\neq 0$}
\begin{figure}[htb]
\begin{center}
\includegraphics[width=12.5cm]{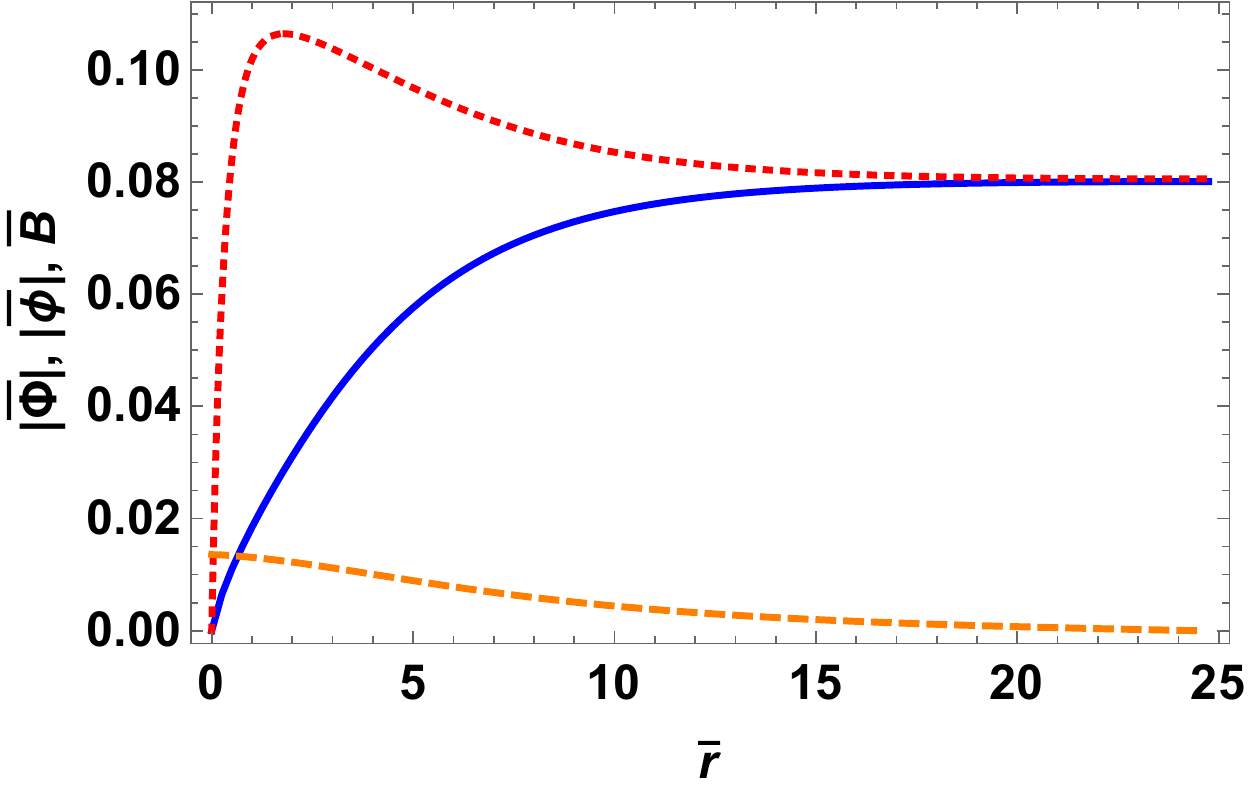}
\end{center}
\caption{The solid (blue) curve represents $\overline{\Phi}$, the dotted (red) curve represents $\overline{\phi}$ and the dashed (orange) curve represents the magnetic field, $\overline{B}$, in a single vortex with $n=1$ for $b=40$, $f=75$.}
\label{fig:vortex1}
\end{figure}

\begin{figure}[htb]
\begin{center}
\includegraphics[width=12.5cm]{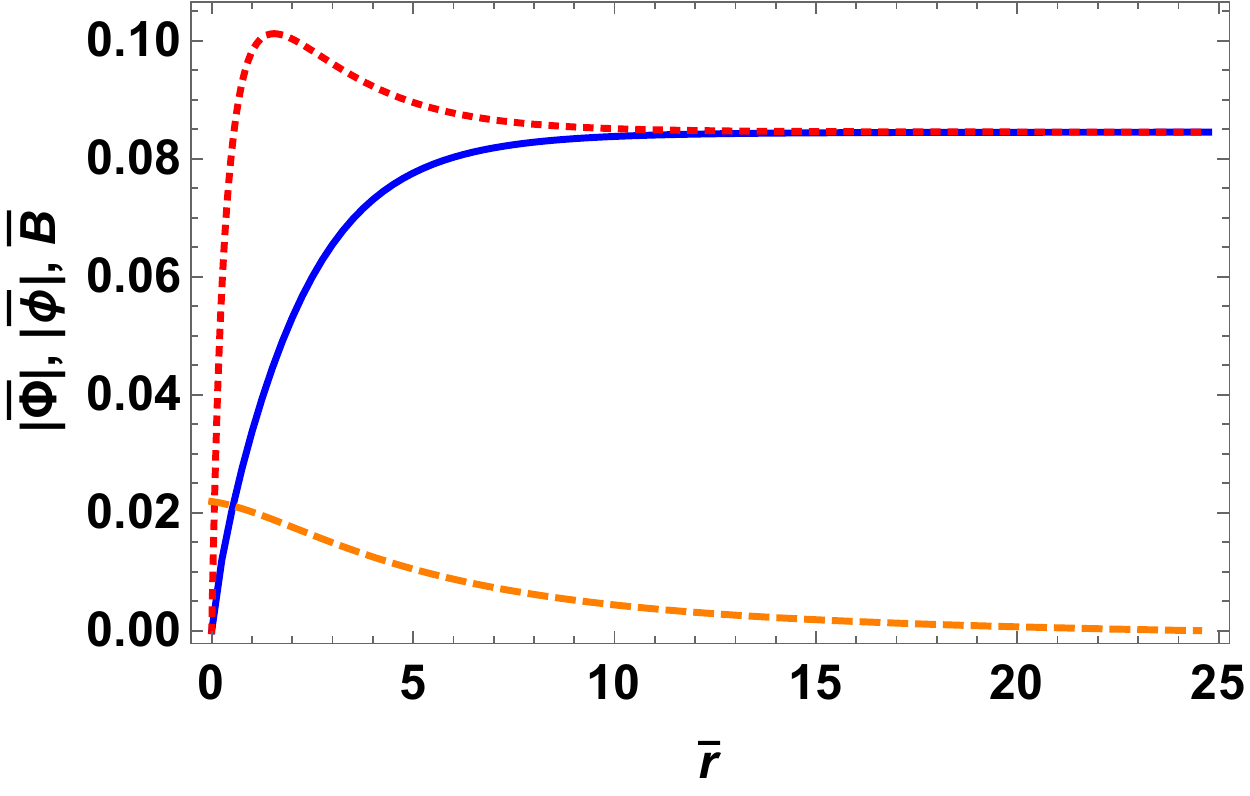}
\end{center}
\caption{The solid (blue) curve represents $\overline{\Phi}$, the dotted (red) curve represents $\overline{\phi}$ and the dashed (orange) curve represents the magnetic field, $\overline{B}$, in a single vortex with $n=1$ for $b=40$, $f=60$.}
\label{fig:vortex2}
\end{figure}

\begin{figure}[htb]
\begin{center}
\includegraphics[width=12.5cm]{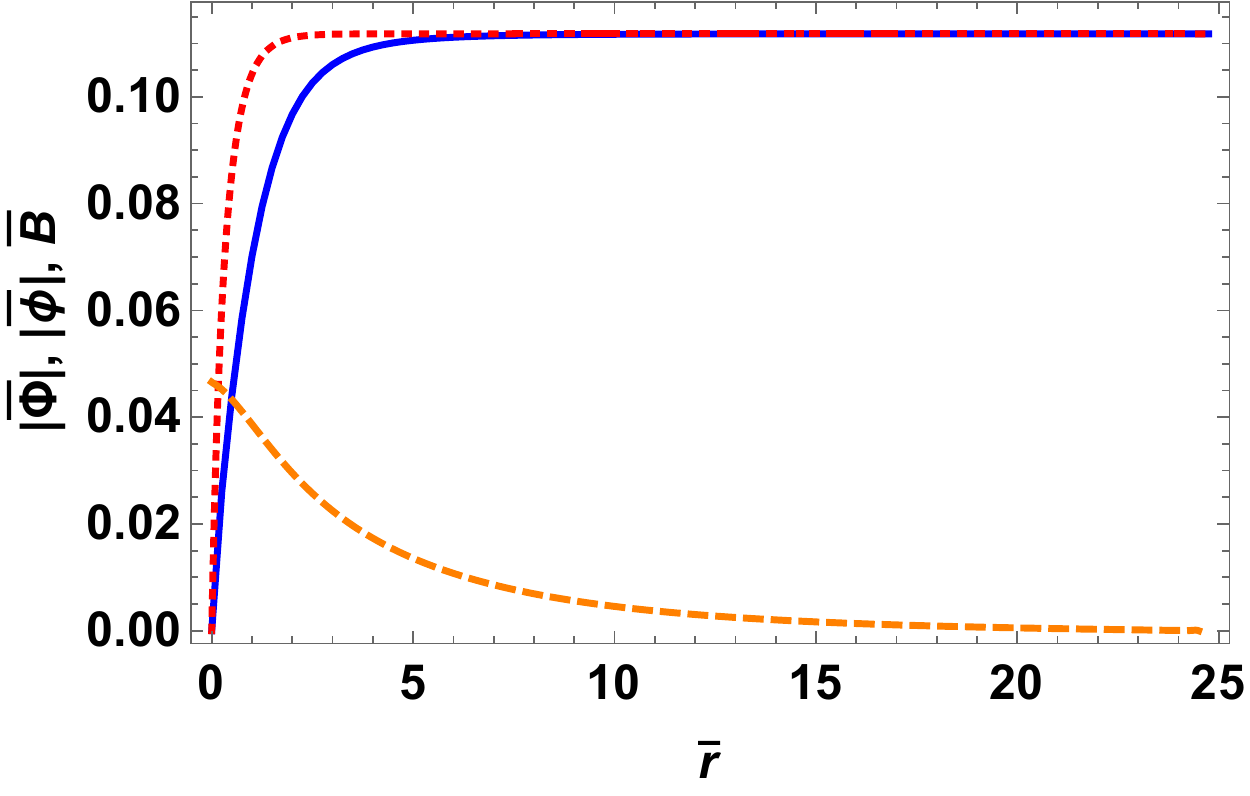}
\end{center}
\caption{The solid (blue) curve represents $\overline{\Phi}$, the dotted (red) curve represents $\overline{\phi}$ and the dashed (orange) curve represents the magnetic field, $\overline{B}$, in a single vortex with $n=1$ for for $b=40$, $f=0$.}
\label{fig:vortex3}
\end{figure}

\begin{figure}[htb]
\begin{center}
\includegraphics[width=12.5cm]{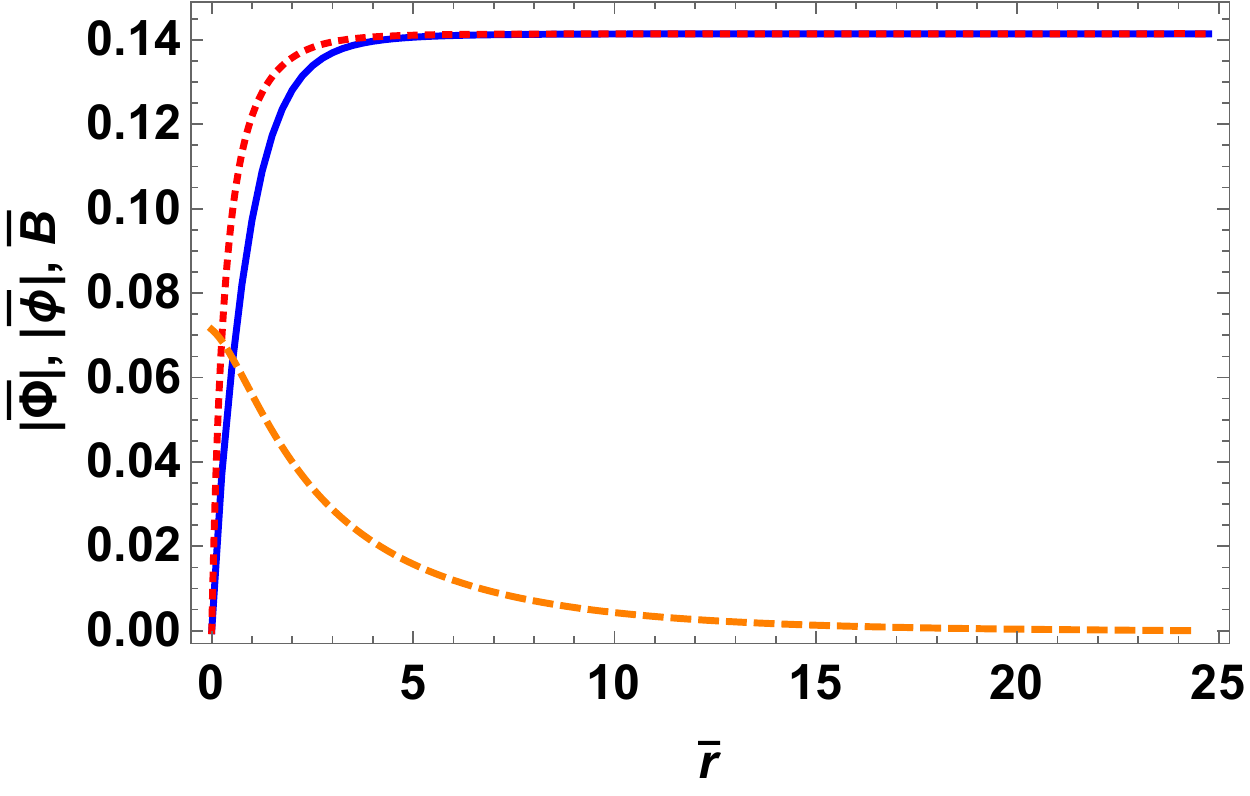}
\end{center}
\caption{The solid (blue) curve represents $\overline{\Phi}$, the dotted (red) curve represents $\overline{\phi}$ and the dashed (orange) curve represents the magnetic field, $\overline{B}$, in a single vortex with $n=1$ for $b=40$, $f=-30$.}
\label{fig:vortex4}
\end{figure}

\begin{figure}[htb]
\begin{center}
\includegraphics[width=12.5cm]{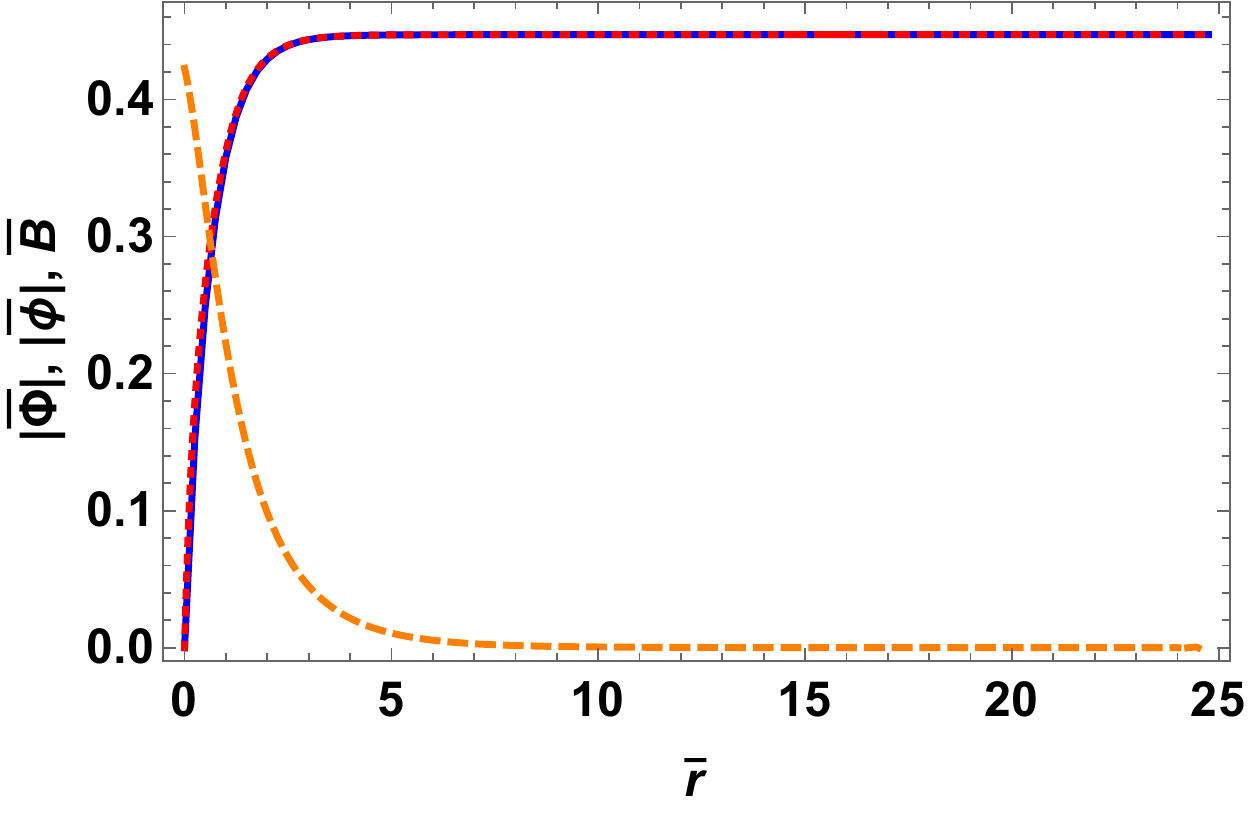}
\end{center}
\caption{The solid (blue) curve represents $\overline{\Phi}$, the dotted (red) curve represents $\overline{\phi}$ and the dashed (orange) curve represents the magnetic field, $\overline{B}$, in a single vortex with $n=1$ for $b=40$, $f=-75$.}
\label{fig:vortex5}
\end{figure}

Here we consider numerically generated vortex solutions assuming $\overline{c}\equiv\frac{c}{a}=1$ and $b=d$ such that the vevs, $v_{\Phi}$ and $v_{\phi}$ are identical. We choose $b=40$ and vary $f$ between $-75$ and $75$. $f$ controls the strength and nature (attractive, repulsive or neither) of the quartic interaction between the $\Phi$ and $\phi$ fields. Figs. \ref{fig:vortex1}, \ref{fig:vortex2}, \ref{fig:vortex3}, \ref{fig:vortex4} and \ref{fig:vortex5} show the resulting vortex structure. 

%uncomment the following lines to place a figure
%\begin{figure}[htb]
%\centerline{%
%\includegraphics[width=12.5cm]{Fig1}}
%\caption{Plot of ...}
%\label{Fig:F2H}
%\end{figure}

For $f>0$, the interaction between the charged scalar and the neutral scalar fields is repulsive and the resulting solutions are such that the scalar fields approach their vevs from opposite sides as shown in Figs.~\ref{fig:vortex1} and \ref{fig:vortex2}. For the choice of parameters in Fig.~\ref{fig:vortex1}, $v_{\Phi}=v_{\phi}=\frac{1}{\sqrt{155}}$ and $\lambda_{-}<\alpha<\lambda_{+}$, suggesting that the behavior asymptotically far from the vortex center is controlled by $\lambda_{-}$. Furthermore, $s_{-}$ is negative, which is consistent with the fact that two scalar fields approach their vevs from opposite directions.

In Fig.~\ref{fig:vortex2}, $v_{\Phi}=v_{\phi}=\frac{1}{2\sqrt{35}}$ and $\alpha<\lambda_{-}<\lambda_{+}$. Therefore, $\alpha$ determines the asymptotic behavior of both the scalar fields. From Eq.~\ref{gandalpha}, we note that $\sign(g_{\Phi})=-\sign(g_{\phi})$, which is consistent with the fact that $\Phi$ and $\phi$ approach $v_{\Phi}$ and $v_{\phi}$ respectively from opposite sides. See Eq.~\ref{fullsolution}.
%%%
%%%

Fig.~\ref{fig:vortex3} shows the result for $f=0$ in which case $v_{\Phi}=v_{\phi}=\frac{1}{4\sqrt{5}}$. In this case, $g_{\phi}=0$, $g_{\Phi}\neq 0$ and $\alpha<\lambda_{+}=\lambda_{-}$, which is the asymptotic behavior of $\Phi$ and $\phi$ are different. The asymptotic behavior of $\Phi$ is controlled by $\alpha$ and that of $\phi$ by $\lambda_{+}$.
%%%
%%%

For $f<0$, the interaction between the scalar fields is attractive and the resulting vortex solutions approach the vevs from the same direction. In Fig.~\ref{fig:vortex4} the vevs are $v_{\Phi}=v_{\phi}=\frac{1}{5\sqrt{2}}$ and in Fig.~\ref{fig:vortex5} $v_{\Phi}=v_{\phi}=\frac{1}{\sqrt{5}}$. In both cases, $\sign(g_{\Phi})=\sign(g_{\phi})$ consistent with numerical results that show the scalar fields approaching their vevs from the same side.
%%%
%%%
%%%
\subsection{$v_{\Phi}\neq 0,\ v_{\phi}=0$}
\begin{figure}[htb]
\begin{center}
\includegraphics[width=12.5cm]{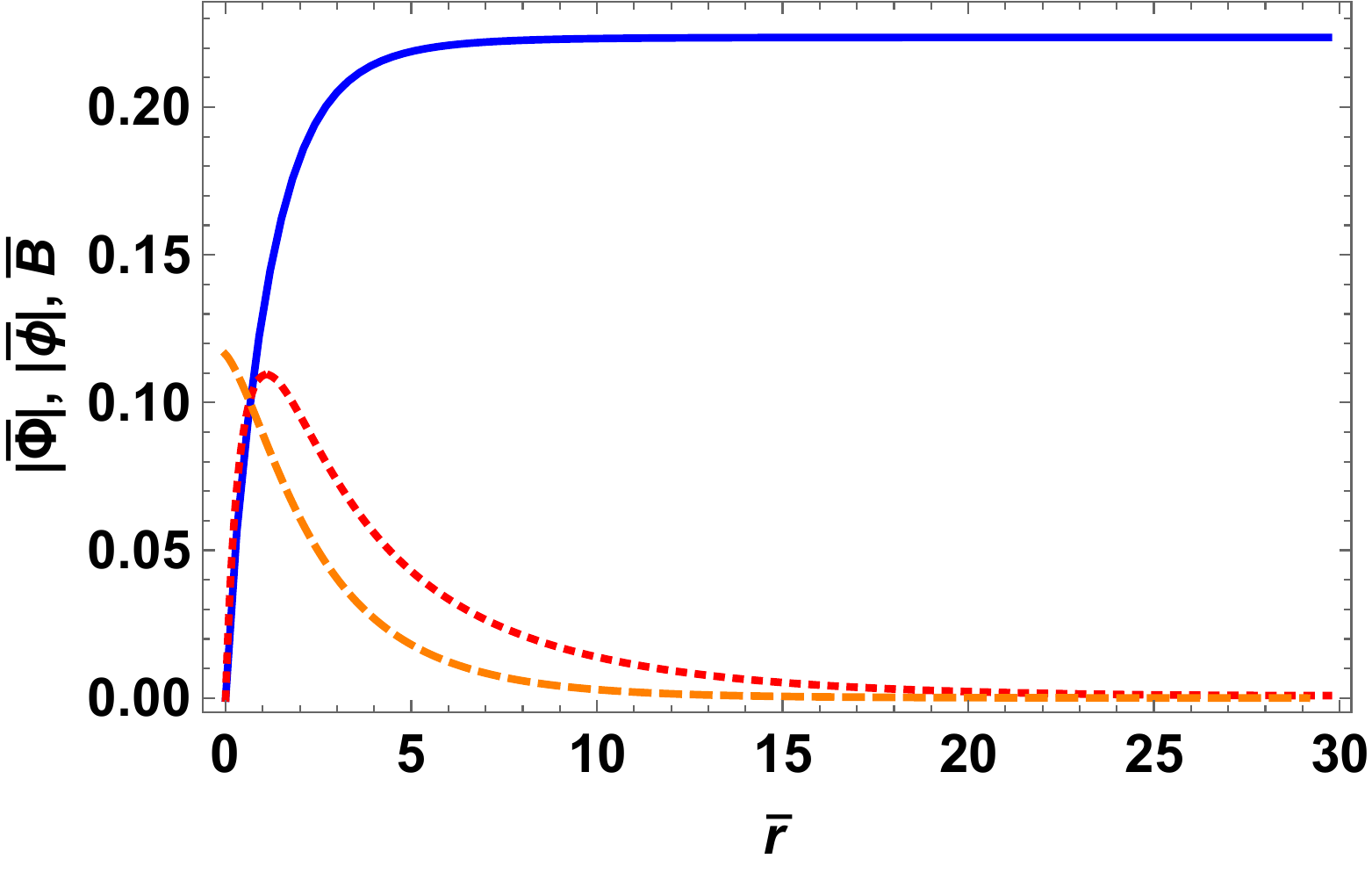}
\end{center}
\caption{The solid (blue) curve represents $\overline{\Phi}$, the dotted (red) curve represents $\overline{\phi}$ and the dashed (orange) curve represents the magnetic field, $\overline{B}$, in a single vortex with $n=1$ for $b=10$, $\overline{c}=\frac{c}{a}=\frac{8}{10}$, $d=b$, $f=16.25$.}
\label{vortex6}
\end{figure}

\begin{figure}[htb]
\begin{center}
\includegraphics[width=12.5cm]{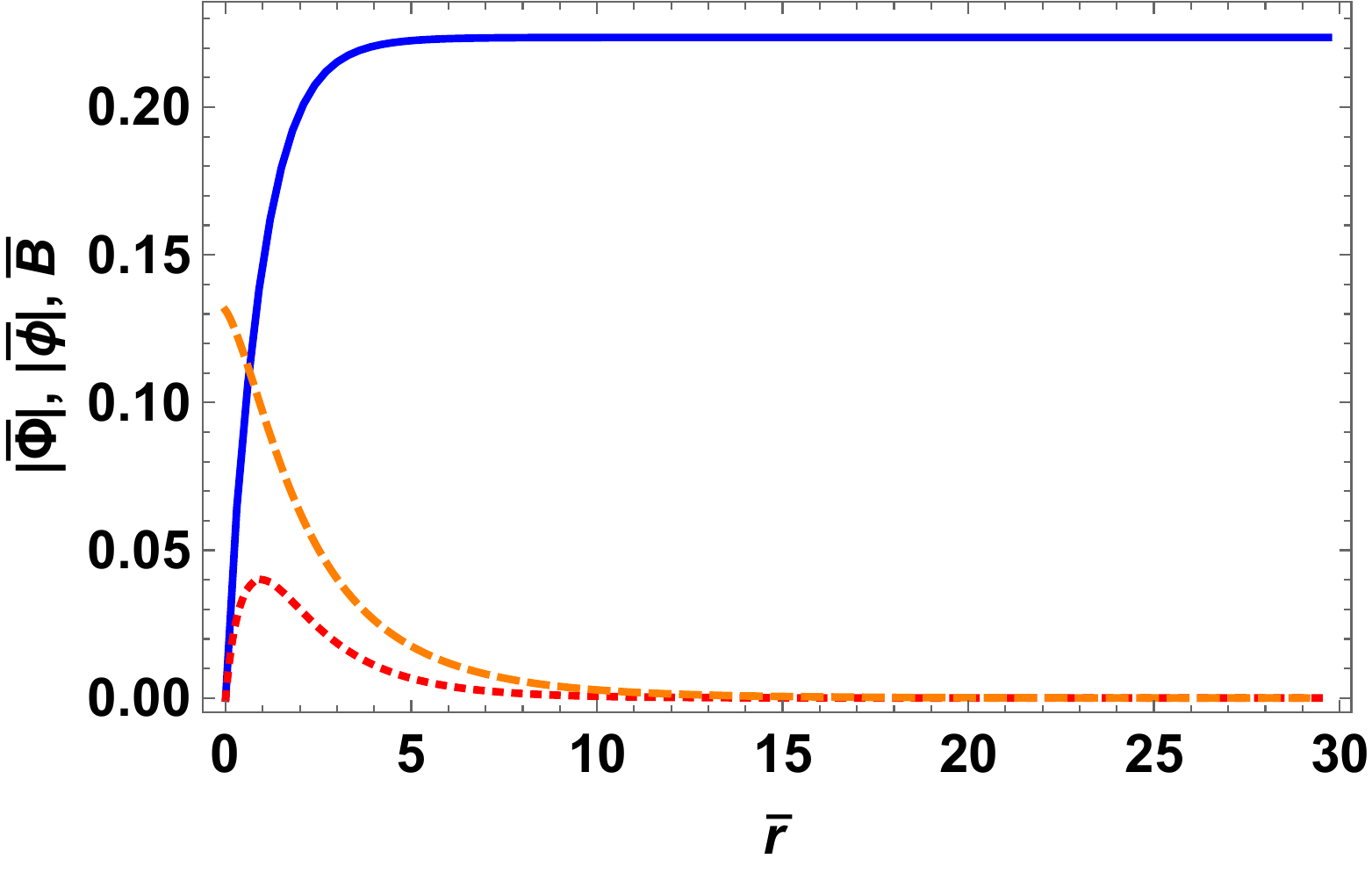}
\end{center}
\caption{The solid (blue) curve represents $\overline{\Phi}$, the dotted (red) curve represents $\overline{\phi}$ and the dashed (orange) curve represents the magnetic field, $\overline{B}$, in a single vortex with $n=1$ for $b=10$, $\overline{c}=\frac{c}{a}=\frac{8}{10}$, $d=b$, $f=17.8$.}
\label{vortex7}
\end{figure}

\begin{figure}[htb]
\begin{center}
\includegraphics[width=12.5cm]{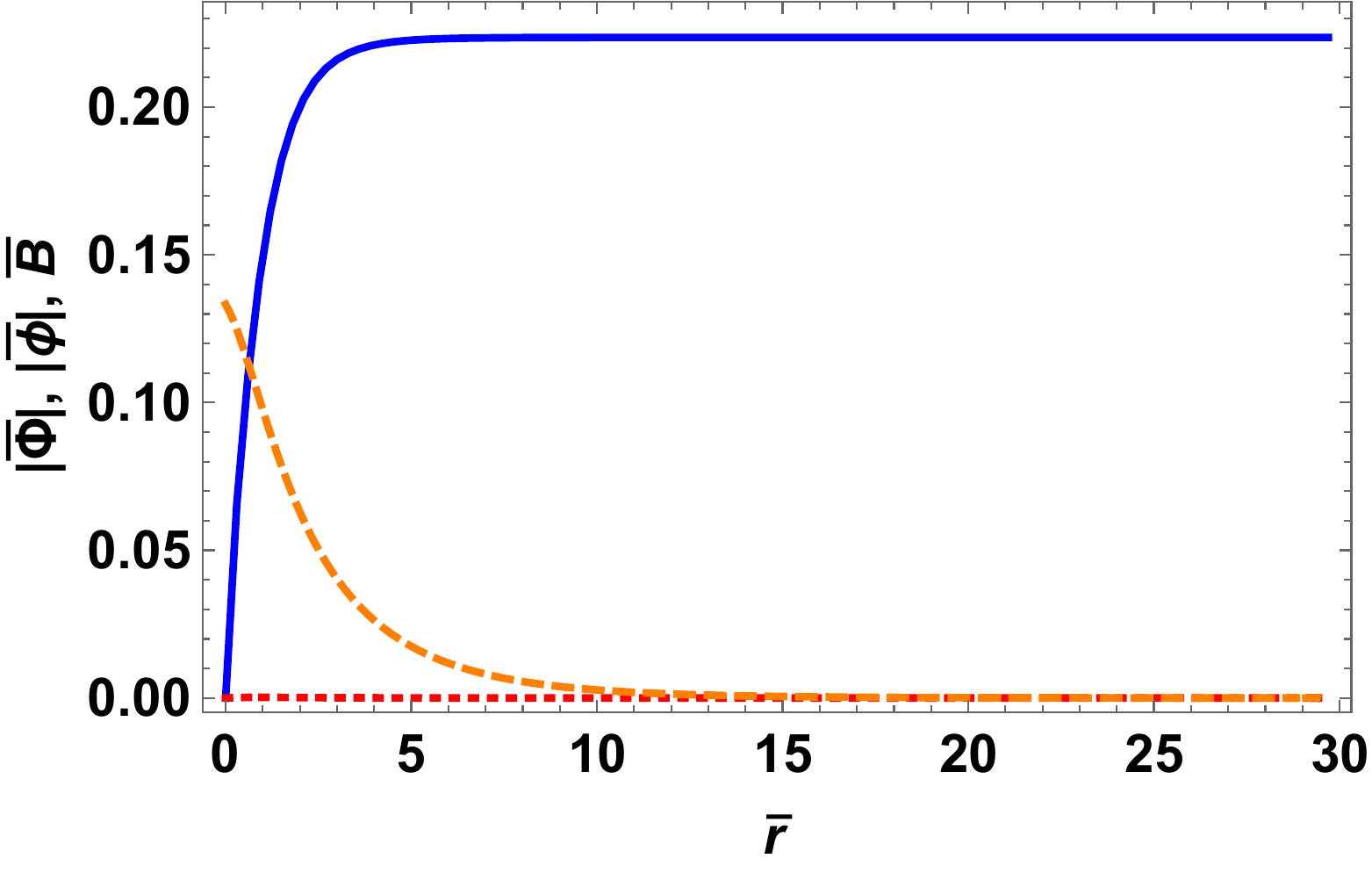}
\end{center}
\caption{The solid (blue) curve represents $\overline{\Phi}$, the dotted (red) curve represents $\overline{\phi}$ and the dashed (orange) curve represents the magnetic field, $\overline{B}$, in a single vortex with $n=1$ for $b=10$, $\overline{c}=\frac{c}{a}=\frac{8}{10}$, $d=b$, $f=18.1$.}
\label{vortex8}
\end{figure}

\begin{figure}[htb]
\begin{center}
\includegraphics[width=12.5cm]{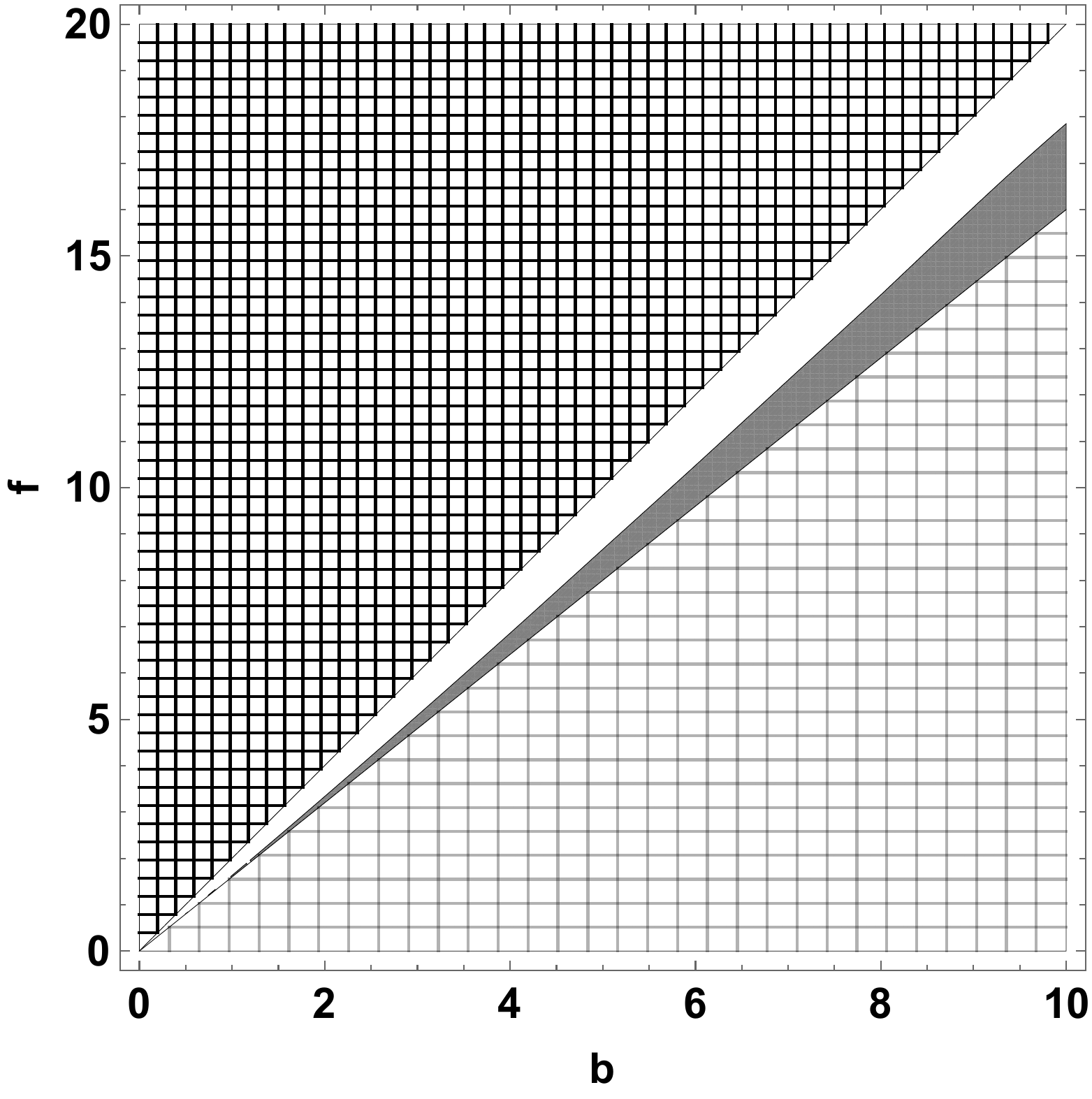}
\end{center}
\caption{``Phase diagram" for $\bar{c}\equiv\frac{c}{a}=\frac{8}{10}$ and $d=b$. In the gray, solid region the real scalar field condenses in the vortex phase but has a zero vev. The region with the black, fine mesh has no stable minimum since $D\equiv 4bd-f^{2}<0$. The coarse, grey mesh shows the region with $af<2bc$, or $f<\frac{16b}{10}$, where both $\Phi$ and $\phi$ have non-zero vevs. In the white region, real scalar remains uncondensed in the vortex state as in the homogeneous ground state.}
\label{phase1}
\end{figure}

\begin{figure}[htb]
\begin{center}
\includegraphics[width=12.5cm]{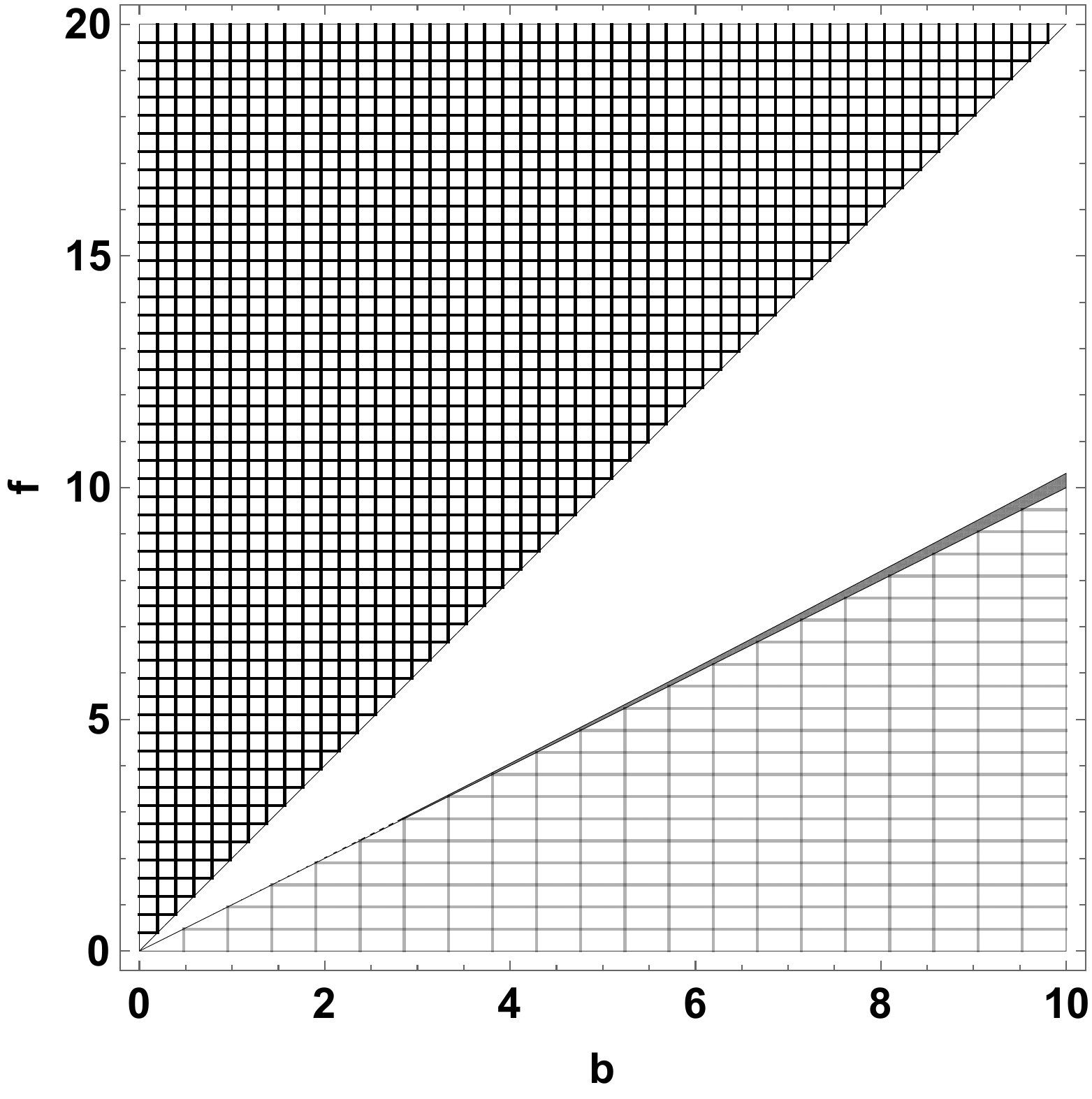}
\end{center}
\caption{``Phase diagram" for $\overline{c}=\frac{c}{a}=\frac{1}{2}$ and $d=b$. In the gray, solid region the real scalar field condenses in the vortex but has a zero vev. The region with the black, fine mesh has no stable minimum since $D\equiv 4bd-f^{2}<0$. The coarse, grey mesh shows the region with $af<2bc$, or $f<b$, where both $\Phi$ and $\phi$ have non-zero vevs. In the solid white region, the real scalar field remains uncondensed in the vortex state as in the homogeneous ground state.}
\label{phase2}
\end{figure}

Finally, we consider the extended Abelian Higgs model with the parameters chosen such that the vev of $\phi$ is zero, $v_{\phi}=0$, but that of $\Phi$ is not, $v_{\Phi}\neq 0$, with particular interest in the possibility of real scalar ($\phi$) condensation within single vortices ($n=1$). We choose $b=d$ but $a\neq c$.
%%%
%%%
%%
Figs.~\ref{vortex6}, \ref{vortex7} and \ref{vortex8} show the resulting vortices for $b=10$ and $\overline{c}=\frac{c}{a}=\frac{8}{10}$ for $f>16$. Note that for $f<16$, $v_{\phi}\neq 0$. The resulting asymptotic behavior of the scalar fields is not given by Eq.~(\ref{fullsolution}). It is easy to see from Eq.~(\ref{ns}) that $n_{2}=n_{3}=0$ and the resulting matrix $N$ of Eq.~(\ref{matrixN}) is diagonal with the non-zero elements being
\begin{equation}
\begin{split}
n_{1}&=2a+12bv_{\Phi}^{2}\\
n_{4}&=2c+2fv_{\Phi}^{2}\ .
\end{split}
\end{equation} 
Noting that $\lambda_{+}=n_{1}$, $\lambda_{-}=n_{4}$ and $g_{\phi}=0$, the resulting solutions for $\delta\Phi$ and $\delta\phi$ are:
\begin{equation}
\begin{split}
\delta\Phi(r)&\approx g_{\Phi}\frac{e^{-\alpha r}}{r}+c_{\Phi}K_{0}(\sqrt{\lambda_{+}}r)\\
\delta\phi(r)&\approx c_{\phi}K_{0}(\sqrt{\lambda_{-}}r)\ .
\end{split}
\end{equation}
It is immediately obviously then that the asymptotics of the $\phi$ field in Figs.~\ref{vortex6}, \ref{vortex7} and \ref{vortex8} are controlled by $\lambda_{-}$ (or $n_{4}$). On the other hand, for the $\Phi$ field asymptotics depends on both $\lambda_{+}$ (or $n_{1}$) and $\alpha$. However, it is straightforward to check that $\alpha<\lambda_{+}$ and as such $\alpha$ determines the asymptotic behavior of $\Phi$.
%%
%%
%%%
In Figs.~\ref{phase1} and \ref{phase2}, we show the resulting ``phase diagram" assuming $\overline{c}=\frac{8}{10}$ and $\overline{c}=\frac{1}{2}$ respectively and $b=d$. The fine black, mesh region shows the region where $D<0$ and as such there is no stable local minimum. In the region depicted by the coarse, gray mesh, both $v_{\phi}\neq 0$ and $v_{\Phi}\neq 0$. In the gray shaded and the solid white regions, $v_{\Phi}\neq 0$ but $v_{\phi}= 0$ with the difference being that in the gray shaded region, real scalars condense in the vortex phase. Finally, note that the region with the real scalar condensed phase is smaller when $\overline{c}=\frac{1}{2}$ compared to the case when $\overline{c}=\frac{8}{10}$.

\section{Conclusion and Future Work}
In this paper, we have constructed an extension of the Abelian Higgs model consisting of an additional real scalar field coupled to the charged scalars via a quartic coupling that preserves the $U(1)$ symmetry of the Abelian Higgs model. The focus of this paper was to study the possibility of real scalar condensation in single vortices with one unit ($n=1$) of magnetic flux (which equals $\frac{2\pi}{e}$) when the vev associated with the real scalar field is non-zero while that associated with the complex scalar is non-zero. We constructed a ``phase diagram" showing the region in parameter space with real scalar condensation. In future work, we will consider the thermodynamics of the extended Abelian Higgs model in the presence of an external magnetic field and consider the structure of Abrikosov lattices and the resulting phase diagram as a function of the external magnetic field and model parameters.

\newpage
\appendix
\section{List of Figures}

\end{document}